\documentclass[aps,prl,twocolumn,showpacs,a4paper,unsortedaddress,amsmath,amssymb,byrevtex]{revtex4}

\usepackage[latin1]{inputenc}
\usepackage{graphicx}
\usepackage{color}
\usepackage{dcolumn}
\usepackage{bm}
\usepackage{hyperref}
\usepackage{times}

\begin{document}

\title{Tuning the electrical conductivity of nanotube-encapsulated metallocene wires}

\author{V\'ictor M. Garc\'{\i}a-Su\'arez$^{1,2}$}
\author{Jaime Ferrer$^2$}
\email{ferrer@condmat.uniovi.es}
\author{Colin J. Lambert$^1$}

\affiliation{
$^1$ Department of Physics, Lancaster University, Lancaster, LA1 4YB, U. K.}
\affiliation{
$^2$ Departamento de F\'{\i}sica, Universidad de Oviedo, 33007 Oviedo, Spain}

\date{\today}

\begin{abstract}
We analyze a new family of carbon nanotube-based molecular wires, formed by encapsulating 
metallocene molecules inside the nanotubes. Our simulations, that are based on a combination
of non-equilibrium Green function techniques and density functional theory, indicate that 
these wires can be engineered to exhibit desirable magnetotransport effects for use in spintronics 
devices. The proposed structures should also be resilient to room-temperature fluctuations, and 
are expected to have a high yield.
\end{abstract}


\maketitle

The ability to fill carbon nanotubes (CNTs) with different compounds \cite{Tsa94} 
has opened a new and growing field of research, where nanowire devices with new electronic 
properties can be produced using simple chemical methods. Examples include encapsulated 
nanowires  made with metallic elements \cite{Guer94}, with fullerenes  that modify 
the local electronic structure of the nanotube \cite{Hor02}, with metallofullerenes that 
modulate the bandgap of semiconductor nanotubes and divide them into multiple quantum 
dots \cite{Lee02}, with organic molecules that allow to produce n- and p-doped 
nanotubes \cite{Jin04,Tak03}, and with iron nanoparticles that facilitate the inclusion of
magnetic properties \cite{Muh03,Jan04}. These experiments demonstrate that it is feasible to
fabricate long and stable chains of single molecules or atomic clusters by placing them in 
the interior of CNTs. They also show how the electronic and structural properties of the CNT 
can be conveniently tailored through the addition of different compounds. Recently, it has 
been demonstrated that metallocenes may be encapsulated inside single-walled CNTs 
with diameters of about 1 nanometer \cite{Nic05}.

Bis-cyclopentadienyl TM (TMCp$_2$) is a metallocene composed of a transition metal atom TM, 
sandwiched between two aromatic rings made of five carbon and five hydrogen atoms each. It may be formed
with the six elements in the middle of the 3d-row, V, Cr, Mn, Fe, Co, Ni, as well as with those
in the column of iron, Ru and Os. All TMCp$_2$ therefore share the same shape, which we depict 
in Fig. 1. Their electronic structure is also very similar, with differences that can be 
traced to the diverse electronic filling of the d-shell of the TM atom. 
The crystal field associated with the benzene rings lifts the degeneracy of the d-shell into a 
triplet that lies lower in energy (d$_{x^2-y^2}$, d$_{xy}$ and d$_{3z^2-r^2}$) and a doublet 
(d$_{xz}$ and d$_{yz}$). The levels in the doublet are split into two peaks due to their 
bonding with the carbon atoms. In addition, there is an exchange field that breaks the spin 
degeneracy to accommodate the magnetism predicted by Hund's first rule. Accordingly, 
VCp$_2$, CrCp$_2$, MnCp$_2$, FeCp$_2$, CoCp$_2$ and NiCp$_2$ have magnetic moments equal to 
3, 2, 1, 0, 1 and 2 Bohr magnetons, respectively. FeCp$_2$ is paramagnetic since its core atom 
has a 
closed-shell electronic structure. The other metallocenes display a sort of mirror symmetry 
about FeCp$_2$. To illustrate these points, we plot in Fig. 2
the density of states projected onto the d-shell of CoCp$_2$.

Such features are highly advantageous, since by placing metallocenes in the interior of CNTs, 
one may fabricate chains of molecules, whose electrical and magnetic properties 
can be tailored at will. We view the CNT as an external coating, whose conducting 
properties can also be modified by changing its chirality from armchair to zigzag. 
Such a coating also provides a strongly enhanced chemical stability to the atomic chain
and isolates it mechanically and, to a point, electrically. 
The simulations that we present here show that the formation energy of these wires is strongly 
exothermic for CNTs of radii between 4.5 and 7 \AA. We therefore argue that these nanometric 
analogues of conventional electrical wires can readily be fabricated. 
A note of caution should be added however since fabrication and sorting of CNTs of a given 
chirality is currently a very difficult task.

In our search for metallocene nanowires, we have performed an extensive series of ab-initio 
simulations of TMCp$_2$ inside CNTs of several chiralities. We have calculated their
electronic and structural properties using the density functional theory code 
SIESTA \cite{Sol02}, that uses pseudopotentials, and expands the wave functions of 
valence electrons by linear combinations of numerical atomic orbitals.
To be specific, we have used a double-zeta basis set to describe hydrogen and carbon atoms,
and a double-zeta polarized basis set for the 3d metal atom. We have also used a number of 
$k$-points ranging between 20 and 100 to model the periodicity along the axis of the nanotube 
and the generalized gradient approximation to the exchange and correlation functional, 
as parametrized by 
Perdew, Burke and Ernzerhof \cite{Per96,correlations}. We have relaxed the atomic coordinates of all 
atoms in the unit cell. 
Finally, we have calculated the electrical conductance of the devices 
with the non-equilibrium Green functions \cite{Dat95,Tran02} code SMEAGOL \cite{Roc06}, 
that has been successfully applied to the study of magnetorresistive effects in organic 
molecules and conductance quantization in atomic constrictions \cite{Roc05,Gar05b}.

\begin{figure}
\includegraphics[width=7cm]{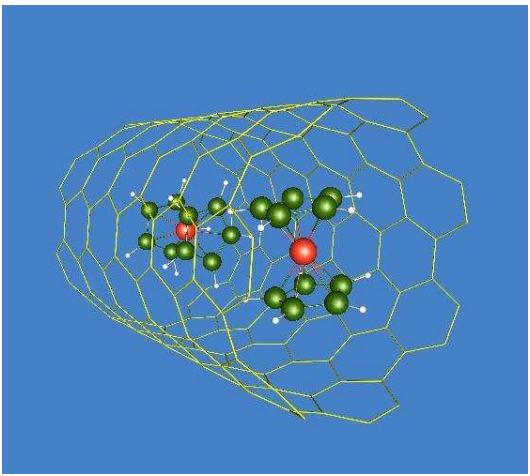}
\caption{
(Color online) Schematic view of a TMCp$_2$@3(7,7) nanotube whith two encapsulated metallocenes,
whose atoms have been highlighted in red, green and white colours (TM, C and 
H atoms, respectively). The axis of the farthest (closest) metallocene is aligned parallel 
(perpendicular) to the axis of the nanotube}
\end{figure}

As a first step, we have simulated single-wall CNTs of different chiralities as well as 
isolated chains of TMCp$_2$. We find lattice constants of 7.88, 7.99, 7.63, 7.55, 7.93 and 
8.20 \AA~ for TM = V, Cr, Mn, Fe, Co and Ni, respectively. Metallocene chains can therefore be 
accommodated rather well inside armchair $(n,n)$ CNTs, whose theoretical lattice constant is 
2.46 \AA, with one molecule every three unit cells of the nanotube. They can also be placed 
inside zigzag $(n,0)$ CNTs, that have a lattice constant of 4.32 \AA, with one molecule every 
two unit cells. Other chiralities may also be chosen for the CNT. For instance, (15,5) and 
(12,4) CNTs can also host metallocene chains (lattice constant of 15.59 \AA, in both cases), 
with two molecules per unit cell.

To test the stability of the encapsulated chains, we have simulated VCp$_2$, FeCp$_2$ and 
CoCp$_2$ chains placed inside metallic $(n,n)$ CNTs with $n$ ranging from 6 to 10, which 
correspond to radii between 3.39 \AA~ and 6.78 \AA, respectively. The structural periodic 
unit of the simulations comprises $N = 2$, 3 or 4 unit cells of the nanotube and a single 
metallocene, and is denoted as TMCp$_2$@$N(n,n)$. In addition, the axis of the metallocene 
has been oriented either parallel or perpendicular to the axis of the nanotube, as shown 
in Figs. 1 (a) and (b). We have calculated the binding energy $E_B$, per structural unit 
cell of the whole nanowire, defined as
\begin{equation}
E_B=E_{{\tt TMCp_2@}N(n,n)}-NE_{(n,n)}-E_{\tt TM{Cp_2}}
\end{equation}
as well as the charge transferred from the metallocene to the CNT $\Delta Q$, and its 
magnetic moment {\it m}, as a function of the radii of the armchair CNTs. Fig. 3 shows 
the results for parallel and perpendicular CoCp$_2$@$N(n,n)$ wires, with $N = 3$ and 4.
We note that CoCp$_2$@$2(n,n)$ is energetically unfavorable since the metallocene molecules
are too compressed, that CoCp$_2$@$3(n,n)$ accommodates rather well the metallocene chain, and 
that the wave functions of the metallocene molecules have negligible overlaps in CoCp$_2$@$4(n,n)$.

\begin{figure}
\includegraphics[width=7cm,height=7cm]{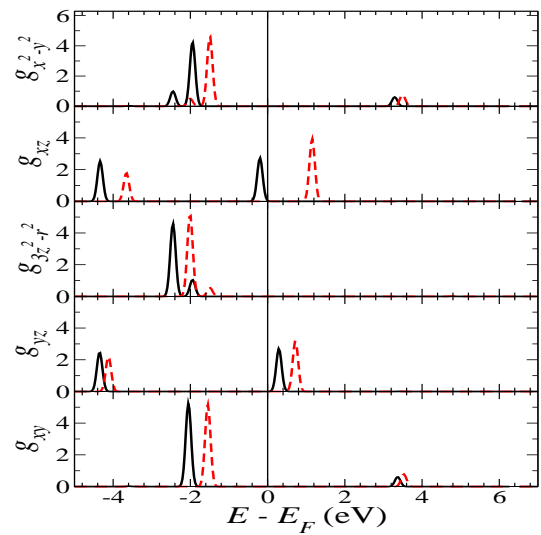}
\caption{(Color online) Densities of states $g_i$ of CoCp$_2$, projected onto the d-shell orbitals of the 
cobalt atom for spin up (black solid line) and down (red dashed line).
The index $i$ denotes the orbital flavor, $i=d_{x²-y²}, d_{xz}, d_{3z²-r²}, d_{yz}, d_{xy}$.}
\end{figure}

Fig. 3 shows that parallel CoCp$_2$@$N(n,n)$ nanowires are energetically favorable for 
radii equal or larger than about 4.5 \AA. The most stable configuration is parallel
CoCp$_2$@$N$(7,7), that corresponds to 
a radius of 4.75 \AA, and has a binding energy as large as 0.85 eV. This result is in excellent 
agreement with the experimental data of Li and coworkers \cite{Nic05}, who found a preferred 
radius of 4.67 \AA. The 
small energy differences between $N = 3$ and $N = 4$ are due to the formation energy of the 
chain. 

We now examine the motion of single cobaltocenes along the nanotube axis. We note that large
energy barriers will hinder the fabrication of chains with no faults, while if the barriers are
too small, metallocenes will move almost freely. This will render the nanowires useless for 
operation at room temperature. We have therefore simulated several configurations of parallel 
CoCp$_2$@$4(n,n)$, where the cobaltocene is placed at different positions along the nanotube axis.
We estimate the energy barrier for motion of the cobaltocene $E_M$ as the energy 
$E_{{\tt CoCp_2@}4(n,n)}$ of each configuration, referred to the energy of the most 
stable position. 
Thin nanowires have energy barriers of the order of several eV. On the contrary, CoCp$_2$@4(7,7) 
has an energy barrier of only 12 meV, which along with its large binding energy, makes this 
nanotube a promising candidate for encapsulating cobaltocenes. For larger radii, $E_M$ decreases 
to values smaller than 3 meV, which allows the molecules to move almost freely along the nanotube,
and eventually leave it \cite{Nic05}.

\begin{figure}
\includegraphics[width=6cm,height=7cm]{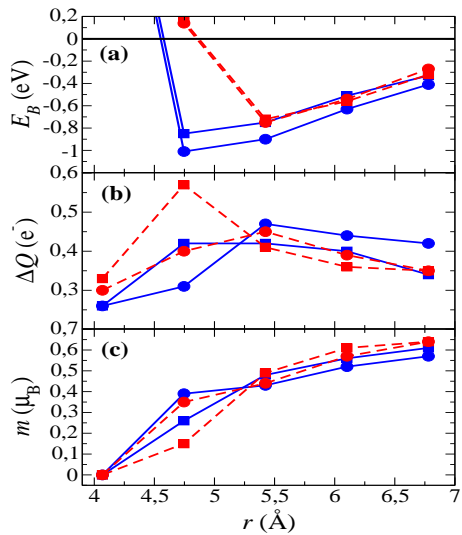}
\caption{(Color online)
(a) Binding energy $E_B$, (b) charge transferred to the nanotube $\Delta Q$, and
(c) magnetic moment $m$, of CoCp$_2$@$N(n,n)$ wires, where $n$ varies from 6 to 10.
Blue solid and red dashed lines correspond to parallel and perpendicular CoCp$_2$@$N(n,n)$, 
respectively. Circles and squares correspond to $N = 3$ and $4$, respectively.}
\end{figure}

To stablish wether or not metallocenes remain perfectly aligned at room temperature, we discuss
now how easy is to rotate a cobaltocene from a parallel to a perpendicular configuration. 
We first note that perpendicular cobaltocenes only fit inside 
CNTs with $n$ equal or larger than 8. We find that the energy cost to rotate the cobaltocene 
for CoCp$_2$@3(8,8) nanowires is as large as 0.15 eV, and therefore they remain aligned at room
temperature. In contrast, metallocenes would be missaligned at room temperature in CoCp$_2$@4(8,8) nanowires, 
due to the negligible overlap of metallocene wave functions that leads to a energy difference 
between both orientations of only 30 meV. 

A Mulliken-population estimate of the total amount of charge transferred from the cobaltocene to 
the nanotube due to formation of the chemical bond, reveals that the nanotube is always 
n-doped as seen experimentally \cite{Nic05} and does not acquire any spin polarization. 
The amount of charge transferred is of
about 0.3 to 0.4 e$^-$, depending on the configuration, as shown in Fig. 3. The values of 
the magnetic moment are considerably smaller than those of the isolated molecule 
($1.0\,{\tt \mu_B}$), due to the charge transferred from the cobaltocene to the nanotube \cite{Jin04}. 

This charge transfer is also evident in the band 
structures shown in Fig. 4, where we plot the spin up and down electronic bands of 
parallel TMCp$_2$@3(7,7), with TM = V, Fe, Co. We note that the relative position of the 
CNT and the transition metal 3d bands indicate the amount of charge transferred and therefore 
the strength of 
the chemical bond. The iron metallocene shows three 3d bands at the Fermi energy $E_F$, that 
are essentially unsplit and dispersionless whereas the CNT bands show a large 
dispersion. This implies that the current flowing through these nanowires must be carried only 
by the external coating. Something similar happens for the vanadium wire, where the bands are 
now exchange-split, but still dispersionless and therefore should not conduct. In contrast, 
the 3d bands of the cobaltocene nanowire are exchange split and dispersive and should carry 
a spin-polarized current. 

\begin{figure}
\includegraphics[width=8cm]{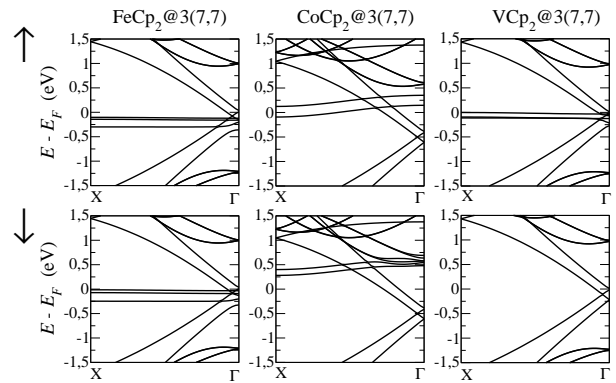}
\caption{
Band structure of parallel TMCp$_2$@$3(7,7)$, for spin up and down electrons, 
where TM = Fe, Co and V. 
The perpendicular case is similar but now also the cobaltocene bands are flat.
Top (bottom) panels are for spin up (down) electrons.}
\end{figure}

The above discussion suggests that parallel CoCp$_2$@3(7,7) and CoCp$_2$@3(8,8) are
the best candidates for spintronics devices and applications. On the one hand, these wires
are the most stable devices from a structural point of view. On the other, their 
band structure shows the presence of a spin-polarized transport channel at the Fermi energy, 
which can give rise to a magnetoresistive effect. 

In order 
to test this possibility we have simulated devices comprised of two leads that sandwich a
scattering region, and computed their conducting properties. The leads contain one unit cell 
of CoCp$_2$@3(7,7) with spin up. 
The scattering region contains also one unit cell of 
CoCp$_2$@3(7,7) with spin up or down. This makes a total of nine nanotube unit cells and 
315 atoms for the simulation. We denote each device by the spin ($u$ or $d$) of the metallocene 
in the scattering region.  Flipping such spin from up to down costs an energy of 0.05 eV, 
which implies that the ferromagnetic configuration can be stabilized at room temperature but that
a relatively small magnetic field may switch it. 
We note that we have not included in our simulations contacts between the devices and external electrodes,
that should decrease to some extent their conducting and magnetorresistive behavior.

The $u$ configuration, in which the spin of the three cobaltocenes points upwards, corresponds
to a perfect ferromagnetic chain, where there is no scattering. In this case the transmission 
coefficient is determined from the band structure and is equal to the number of open scattering 
channels. Fig. 5 shows that there are three channels for spin up in a region around the Fermi 
energy that extends from -0.1 eV up to 0.4 eV. Two of them correspond to the nanotube and one to the cobaltocene. This excludes a tiny segment, where both cobaltocene bands contribute. 
On the other hand, there are only two channels at the Fermi energy for spin down, that 
correspond to the nanotube, while the cobaltocene channels appear 0.3 eV above $E_F$.

In order to study the change of resistance as a function of the magnetic configuration we have 
also calculated the conductance of the $d$ configuration. In this case, the cobaltocene channel
has moved to energies way above $E_F$ due to the exchange splitting, whereas the transmission 
for spin down is unaffected. As a consequence, the 
difference in conductance between $u$ and the $d$ configurations is 
$0.9\,e^2/h$. Defining 
the magnetoresistance ratio as $\delta G=(G_{u}-G_{d})/G_{d}$, the predicted ratio is
$\delta G=20\%$.

\begin{figure}
\includegraphics[width=7cm]{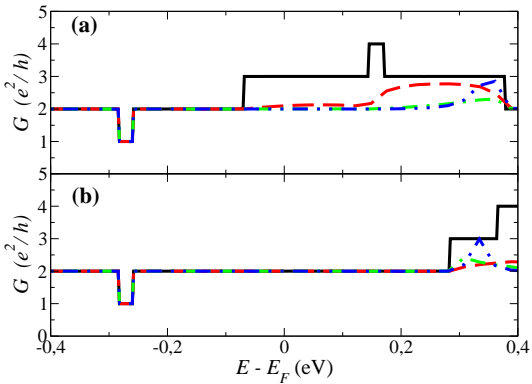}
\caption{(Color online)
Conductance of the CoCp$_2$@3(7,7) magnetorresistive devices for (a) spin up and 
(b) spin down electrons.
$u$, $d$, $dd$ and $ddd$ conductances are plotted with black solid, red dashed, green 
dash-dotted and blue dash-double-dotted lines, respectively.}
\end{figure}

Since the spin flipping of only one cobaltocene may be difficult to achieve in practice, we have 
simulated two other configurations that have two and three CoCp$_2$@3(7,7) unit cells in the scattering region, with all its metallocenes pointing downwards. These configurations, that
have a total of 420 and 525 atoms, are denoted by $dd$ and $ddd$, respectively. As shown in 
Fig. 5, the resulting transmission for spin up channels around the Fermi energy decreases slightly. This yields 
an additional increase in the magnetorresistance ratio to $25\%$.
We have also simulated devices where the spin of the leads are either aligned or 
anti-aligned, but where there is no cobaltocene in the scattering region. In this case the magnetorresistance ratio vanishes. This confirms the fact that the spin 
polarized channel is present only when the cobaltocenes form an unbroken conducting chain, with 
dispersive bands.

Finally we have examined the effect of the conducting properties of the CNT coating.  With a 
view to 
improving the magnetoresistance ratio, it is desirable to increase the fraction of the current 
passing through the metallocene chain and so instead of using armchair CNTs, which are always 
metallic, the question arises of whether we could engineer an insulating coating by choosing 
non-conducting zigzag or chiral CNTs. Our calculations show that the charge transferred from the 
metallocene to the CNT due to the chemical bonding brings the CNT valence bands down in energy.
The CNT coating becomes either conducting or semimetallic, as we show in Fig. 6 for 
CoCp$_2$@$2(11,0)$), VCp$_2$@$2(11,0)$ and 2CoCp$_2$@$1(12,4)$.

\begin{figure}
\includegraphics[width=8cm]{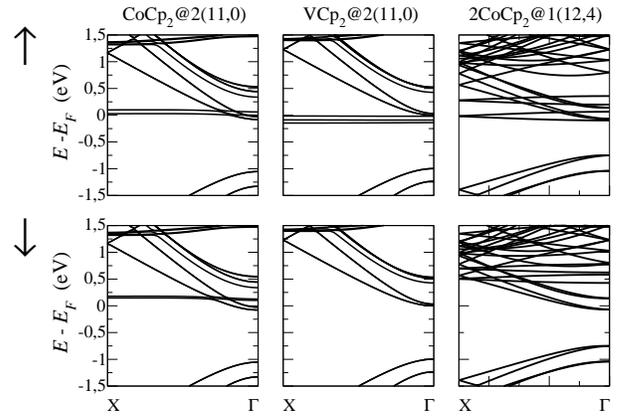}
\caption{
Band structure of parallel CoCp$_2$@$2(11,0)$, VCp$_2$@$2(11,0)$ and 2CoCp$_2$@$1(12,4)$ for
spin up (top panels) and down electrons (bottom panels).}
\end{figure}

In summary, we have demonstrated that electrical nanowires consisting of a core made of a metallocene
chain and a CNT coating may be fabricated, whose magnetic and electrical properties can be 
tailored by the appropriate choice of metallocene molecule and chirality of the CNT. We have performed 
a series of ab-initio simulations of the electronic and structural properties of several TMCp$_2$@$(n,n')$, 
which  agree very well with recent experiments and show that engineering of 
molecular spintronics properties
is possible. 

\begin{acknowledgments}
We acknowledge conversations with John Jefferson and Robin Nicholas, and
financial support from the European Commission, the Spanish Ministerio de
Educaci\'on y Ciencia and the British EPSRC, DTI, Royal Society and NWDA.
\end{acknowledgments}


\begin{thebibliography}{99}
\bibitem{correlations}
We note that our calculations treat electron-electron interactions
at a mean-field level. Still, we believe that the most important
correlation effects that may appear in the proposed devices take the form
of mixed valence behavior, which should be captured at least qualitatively by our
simulations.
\end{thebibliography}
\end{document}